\documentclass[aps,prd,a4paper,superscriptaddress,nofootinbib,showpacs,showkeys,amsfonts,amssymb,amsmath]{revtex4}

\begin{document}

\title{Compact objects from gravitational collapse: an analytical toy model}


\author{Daniele Malafarina} \email{daniele.malafarina@nu.edu.kz}
\affiliation{Department of Physics,
Nazarbayev University, 53 Kabanbay Batyr avenue, 010000 Astana, Kazakhstan}
\author{Pankaj S. Joshi} \email{psj@tifr.res.in}
\affiliation{Tata Institute of Fundamental Research, Homi Bhabha Road,
Colaba, Mumbai 400005, India}

\begin{abstract}
We develop here a procedure to obtain regular static
configurations as resulting from dynamical gravitational collapse
of a massive matter cloud in general relativity.
Under certain general physical assumptions for the collapsing
cloud, we find the class of dynamical models that lead to an
equilibrium configuration. To illustrate this, we provide
a class of perfect fluid collapse models that lead to a
static constant density object as limit.
We suggest that similar models might possibly constitute the
basis for
the description of formation of compact objects in nature.
\end{abstract}


\maketitle

\section{Introduction}
We know today that in nature compact objects such as white
dwarfs and neutron stars do form from gravitational collapse of
massive stars that finally settle to a regular, stationary
configuration. Typically, such a compact source that
results from collapse is supported by nuclear or electromagnetic
forces in terms of internal pressures, but we do know that
gravity plays a crucial role both in the process of collapse
itself as well as in the final equilibrium configuration.
This is true also for larger entities in the universe, such
as galaxies and clusters of galaxies that may have a very massive
compact core at the center.
Therefore it is useful
to investigate the role played by gravity
in the formation of compact gravitating objects.
The general theory of relativity constitutes the
natural framework for such a study whenever the gravitational field
becomes strong (i.e. when the size of the object approaches
its Schwarzschild radius).

In fact, static regular perfect fluid compact sources
in general relativity have been used to model very dense
objects such as neutron stars for a long time now
(see for example
\cite{elasticity} or \cite{Brito}
and references therein). Furthermore a theory
of relativistic elasticity provides the first step towards
a physically meaningful description of the inner
structure of such objects
\cite{Magli}.
Also relativistic models for compact sources of several kinds
have been proposed as description of yet to be found
astrophysical exotic objects. These vary from boson stars (see
\cite{boson} and references therein),
to gravastars (see
\cite{gravastar}),
preon stars (see for example
\cite{preon})
and quark stars (see for example
\cite{quark}).
Even though their existence is not proven the search for
theoretical models that describe exotic compact objects is
very important since it is related to many of the unsolved
problems of modern astrophysics, such as dark matter,
dark energy, the formation of structures in the universe
and the final fate of gravitational collapse of massive stars.

While the astrophysical aspects of such general
relativistic models are very much under discussion and
further investigation is needed, we certainly know that
the physical process that gives rise to neutron stars in
nature is the gravitational collapse of a massive star,
somewhere in the range of 4-40 solar masses.
In such a collapse process that takes place at the very
end of the life-cycle of a massive star, the outer layers
of the star are blasted away in a supernova explosion
while the inner heavy core implodes to form a neutron star,
a very high density object of the radius of the order of 10 kilometers.

As mentioned above, there have been attempts
to model such static massive compact objects within a
relativistic framework, both analytically and via numerical
simulations.
Analytical models are particularly interesting as they provide
insights on the properties that the interior geometry of such objects
must have
(see for example \cite{Martin}).
Also, from the study of such models it is possible to derive important
information on
the emergence of negative pressures and violation of energy conditions
that are expected to happen close to the core when the boundary of the
object approaches or surpasses the event horizon.
This provides some insights on the nature of
the equation of state for extremely dense matter purely from a relativistic
perspective (see for example \cite{MM}).
However,
due to the intrinsic difficulty of Einstein's field equations,
 the general relativistic
description of dynamical collapse itself that settles
to such a final static configuration has not been
explored so far in detail.
Such a process, and especially the late stages of the
gravitational collapse are the phases where the general
relativistic effects should certainly be important, and
in fact these could be the dominant ones which
really rule the final outcome of collapse.
It is therefore very useful to investigate,
within the framework of the gravitation theory, the
scenarios where a dynamical gravitational collapse of a
massive matter cloud would evolve to result into an eventual
regular static configuration. In this direction in
\cite{JMN}
it was shown that large equilibrium static configurations
can be obtained from gravitational collapse, when certain
pressures are present.
In
\cite{JMN2}
the observational features of such exotic compact
objects were studied.
Such studies are very important from the point
of view of astrophysics as they can tell us how to
distinguish an exotic compact object from a black hole.
For example, in the case of gravastars it has been shown that a
realistic model must necessarily have anisotropic pressures
\cite{cattoen}
and that the spectrum of quasinormal modes of such objects is
considerably different from that of a black hole of the
same mass
\cite{gravastar2}.

In the present work we would like to ask and consider
the question if one can obtain general relativistic
collapse that, neglecting the presence of other forces, still
settles to a static regular configuration. We find the answer
to be in the affirmative, and we present here a class
of perfect fluid collapse models tending to an equilibrium
configuration which is described by a well-known static
interior.
The perfect fluid collapse model here, although lacking a
constitutive equation relating pressure and density, can be
considered physically viable since it satisfies all basic
physical requirements (such as energy conditions, regularity,
continuity equation for the matter fields) and develops from
a physically realistic initial state to a well known and physically
viable final state.


In Section 2, we give the general procedure
to generate an equilibrium configuration as the final state
of the gravitational collapse of a perfect fluid. Then
Section 3 illustrates this procedure with a specific application
showing as to how the constant density equilibrium objects
result from collapse. The last section summarizes concluding
remarks which try to bring out the relevance of gravitational
collapse in general relativity towards generating compact
static objects in nature. In the following we make use of geometrical units for which $G=c=1$.

\section{Static configurations from gravitational collapse} \label{collapse}
The most general spherically symmetric spacetime
describing a gravitationally collapsing matter cloud takes
the form,
\begin{equation}
    ds^2=-e^{2\phi}dt^2+\frac{R'^2}{G}dr^2+R^2d\Omega^2 \; ,
\end{equation}
where $\phi$, $R$ and $G$ are functions of the
comoving coordinates $t$ and $r$.
For a matter source made of perfect fluid the energy-momentum
tensor is given by, $T^0_0=-\rho, \; T_1^1=T_2^2=T_3^3=p$.
The density and pressure are then coupled to the metric
functions via Einstein equations, which in
this case can be written as,
\begin{eqnarray}\label{p}
  p &=&-\frac{\dot{F}}{R^2\dot{R}} \; , \\ \label{rho}
  \rho&=&\frac{F'}{R^2R'} \; , \\ \label{phi}
  \phi'&=&-\frac{p'}{\rho+p} \; ,\\ \label{Gdot}
  \dot{G}&=&2\frac{\phi'}{R'}\dot{R}G \; ,
\end{eqnarray}
where $(')$ denotes a derivative with respect to $r$ and $(\dot{})$
denotes derivatives with respect to $t$.

In the above, the Misner-Sharp mass $F(t,r)$ describes
the amount of matter enclosed by the shell labeled by $r$
at any given time $t$, and is given by,
\begin{equation}\label{misner}
F=R(1-G+e^{-2\phi}\dot{R}^2) \; .
\end{equation}
We note that since the system of Einstein equations
has six unknowns, namely $p$,
$\rho$, $\phi$, $G$, $F$ and $R$ in five equations,
there is the freedom to choose one free function.
Once such a choice is made, for example by supplying
the mass profile $F$ for the collapsing cloud, then
the full collapse evolution and the final state of
collapse is determined fully by the Einstein equations
above. On the other hand, equivalently, providing
an equation of state for the collapsing matter that
relates the pressure to the energy density also
closes the system completely.

In order for the collapse model to be physically
realistic we must require a few physical reasonability
conditions. These include absence of shell crossing singularities
(which implies the condition $R'>0$), the weak energy condition
that imposes positivity of the energy density (namely $\rho>0$)
and of the sum of density and pressure ($\rho+p>0$),
regularity of initial data and also absence of trapped
surfaces at the initial time from which the collapse
develops.
Finally the collapsing cloud must in general be
matched to a generalized Vaidya exterior spacetime at
the boundary $R_b(t)=R(r_b,t)$ in the case where there is
a mass inflow or outflow from the star during the collapse phase
\cite{Senovilla}.
In the case where the pressure vanishes at the boundary
and the Misner-Sharp mass within the boundary is conserved,
the matching can be done with an exterior Schwarzschild
spacetime
\cite{matching}.

Here we take without any loss of generality, that
the dependence of $F$ in $t$ goes through $R$ in the form
$F=F(r,R)$, which is always possible whenever $R$ is monotonic
in $t$ (as is the case for any gravitational collapse).
If we want to describe collapse the prescription
$\dot{R}\leq0$ is enough to ensure that the cloud is not
expanding and if $\dot{R}=0$ is reached asymptotically,
this ensures the desired monotonic behaviour. In this
case we can perform a change of coordinates from $(r,t)$
to $(r,R)$, thus considering $t=t(r,R)$ (in the following
$(,r)$ will be used to denote a derivative with respect
to $r$ in the $(r,R)$ coordinates, so that $F'=F_{,r}+F_{,R}R'$).
Using the additional gauge freedom that comes from the
scaling of the model, we set the initial time so that
$R(r,t_i)=r$.

To solve the system of Einstein equations then we
must first of all fix the behaviour of the free function
that is left in the system.
In the present case we choose this to be the mass function
$F(r,R)$. Supplying the mass function then fixes the behaviour
and the future time evolution of the collapsing matter
cloud. Then equations (\ref{p}) and (\ref{rho}) give
$p(r,R)$ and $\rho(r,R)$ as functions of $R$. Integration
of equations (\ref{phi}) and (\ref{Gdot}) then gives
the metric functions $\phi$ and $G$ as:
  \begin{eqnarray}
    \phi(r,R)=-\int_0^r\frac{p'}{\rho+p}d\tilde{r}, \\ \label{G}
    G(r,R)=b(r)e^{2\int_r^R\frac{\phi'}{R'}d\tilde{R}},
  \end{eqnarray}
where $b(r)$, which is usually called the velocity profile
for the collapsing shells, is another free function coming
from the integration and which is related to the kinetic energy
of the infalling matter shells.
The system is then solved once we integrate the
Misner-Sharp mass equation (\ref{misner}), written in the form
  \begin{equation}\label{motion}
    t_{,R}=-\frac{e^{-\phi}}{\sqrt{\frac{F}{R}+G-1}},
  \end{equation}
to obtain the function $t(r,R)$, or equivalently
the physical radius of the cloud $R(r,t)$.

With this, the metric describing the collapsing
spacetime is then given as,
\begin{equation}\label{metric}
    ds^2=-e^{-2\int_0^r\frac{p'}{\rho+p}d\tilde{r}}dt^2+\frac{R'^2}{b(r)
e^{2\int_r^R\frac{\phi'}{R'}d\tilde{R}}}dr^2+R^2d\Omega^2 .
\end{equation}
It is clear of course that in general it might not be possible
to fully integrate the system of Einstein equations, but that
may not always be needed also.

Our main aim here is to construct the dynamical collapse
of a massive cloud that settles to an equilibrium state in which
the pressure balances the gravitational attraction. In order
to achieve this we must choose the mass function and
the velocity profile suitably. In fact in general the dynamics
as implied by the Einstein equations can lead to three
possible final outcomes for a continual gravitational collapse:
Firstly, an indefinite complete collapse (as in the case of a
dust cloud, for example), where all matter falls into the
central spacetime singularity, which may be covered in an
event horizon or may possibly be visible to an external
faraway observer. Secondly, there may be a bouncing behaviour
where the infalling matter shells re-expand after reaching a
minimum radius. Finally, the collapse may settle to obtain a
static final object, in which case we must balance the
velocities and pressure of the matter content of the cloud
in order to obtain the limiting behaviour that lies
between the two earlier outcomes.

The equation of motion (\ref{misner})
can be written as an effective potential for any fixed $r$ as,
\begin{equation}\label{V}
    V(r,R)=-\dot{R}^2=-e^{2\phi}\left(\frac{F}{R}+G-1\right) \; .
\end{equation}
If $\dot{R}<0$ at all times
and $\dot{R}=0$ is not reached even asymptotically,
the collapse then does not halt and
all the matter is crushed into the final central singularity
resulting in the formation of a black hole or a naked singularity
\cite{initial}.
If the final outcome is to be a static, regular configuration
then some conditions must be imposed on the pressure profile so
that the collapse will eventually halt, reaching zero velocity
and acceleration.
The conditions that must be imposed so that the metric
evolves towards an equilibrium configuration therefore are,
\begin{equation}\label{static}
    \dot{R}=\ddot{R}=0 \; ,
\end{equation}
and they are equivalent to $V=V_{,R}=0$.

It is clear of course that a static configuration
cannot be achieved for the dust {\it i.e.} pressureless collapse
models, where $V$ is negative at all times. Nevertheless from
the analysis of the potential when non-zero pressures are
introduced, we see that $V$, as a function of $R$ for any fixed
shell $r$, can have in general two zeros. Then the static
configurations can be obtained from those potentials for which
both the zeroes coincide. This implies that $V$ has a local maximum
for which $V=0$. By linearizing the potential close to the
equilibrium point we can see that the static configurations
can be achieved in the present comoving coordinate system
asymptotically as $t$ grows to infinity.

If the solution of the equation of motion (\ref{motion}),
given by $R(r,t)$, tends asymptotically to an equilibrium solution
$R_e(r)$ such that the conditions given by equations (\ref{static})
are satisfied, then the collapse evolves towards a static
equilibrium configuration.

In order to obtain the equilibrium scenario we
must then choose the free function $F(r,R)$ during collapse
in such a way that the quantities $F$, $\phi$, and $G$ tend
to their equilibrium limits, namely $F(r,R) \rightarrow F_e(r)=F(r,
R_e(r))$ and similarly for $\phi$ and $G$, where $\phi_e(r)$
is given by the Einstein equations and $G_e(r)$ is determined
by the imposition of the static conditions (\ref{static}):
\begin{eqnarray}\label{ve}
  G_{e}(r)&=&1-\frac{F_e}{R_e} \; , \\ \label{Ge}
  (G_{,R})_e &=&G_{,R}(r, R_e(r)) = \frac{F_e}{R_e^2}-\frac{(F_{,R})_e}{R_e} \; .
\end{eqnarray}
Note that the velocity profile $b(r)$ appearing in equation (\ref{G}) has been
absorbed into $G_e(r)$.
The above equations are direct consequence of the
equilibrium conditions (\ref{static}). The key point here
is the following: Given any static equilibrium configuration,
which is characterized by specifying the mass function
$F_e(r)$ of the static system, the free function $F(t,r)$
during the collapse is to be chosen in such a way that
$F(t,r) \to F_e(r)$ in the limit. The class of all such
dynamical collapses as specified by this condition will
then settle to the equilibrium limit.

At the equilibrium we can define a new radial
coordinate from $R_e(r)=z$ and rewrite the metric
functions as,
\begin{equation}
  F_e(r)=\bar{F}(z), \;
  \phi_e(r)=\bar{\phi}(z), \;
  G_e(r)=\bar{G}(z)=1-\frac{\bar{F}}{z}.
\end{equation}
Then from equations (\ref{rho}) and (\ref{phi}) we
recover the Einstein equations for a static interior,
given by,
\begin{equation}
  \rho(z) = \frac{\bar{F}_{,z}}{z^2}, \;
  p_{,z} = -(\rho+p)\bar{\phi}_{,z},
\end{equation}
where the second equation is the well-known
Tolman-Oppenheimer-Volkoff equation. Then the third
static Einstein equation, namely
\begin{equation}
    p=\frac{2\bar{\phi}_{,z}}{z}\bar{G}(z)-\frac{\bar{F}(z)}{z^3} \; ,
\end{equation}
can be obtained from equation (\ref{p}) once we impose
the equilibrium conditions and we make use of
equation (\ref{Gdot}) at equilibrium.

The spacetime geometry given by the metric (\ref{metric})
during collapse, once the equilibrium is reached, reduces to
the familiar static spherically symmetric geometry given by
\begin{equation}\label{metric-eq}
ds^2=-e^{2\bar{\phi}}dt^2+\frac{dz^2}{\bar{G}}+z^2d\Omega^2 \;,
\end{equation}
which can be matched to a Schwarzschild vacuum exterior
at the boundary $z_b=R_e(r_b)$.
We note that the equilibrium configuration again is
fully determined by the choice of one free function,
which we take to be the mass function $\bar F(z)$.
An important point to notice here is that the above metric
in principle need not be necessarily regular at the center
since any singularity that might eventually form is
obtained as the result of continual collapse from a
regular initial data.

Generally, when studying static perfect fluid sources
of the Schwarzschild spacetime, one requires a set of physically
viable conditions to be satisfied
\cite{Lake}.
These are the usual energy conditions, the matching
conditions with exterior Schwarzschild geometry given by
the vanishing of the pressure at the boundary, a monotonically
decreasing behaviour for the energy density and pressure, and
regularity at the center. As we have seen, this last
requirement of regularity at center in the static final
state could in principle be omitted. Finally one wishes
to impose also that the sound speed in the cloud be smaller
than the speed of light, thus requiring $p/\rho<1$.

We note that in general if energy conditions are
satisfied during collapse they will be satisfied by the
equilibrium configuration as well, namely the positivity
of the energy density and sum of density and pressure at the
origin follows from the same condition during collapse.
Further we note that requiring only the weak energy condition
allows for the possibility of some kind of negative
pressures.

\section{Constant density interiors as collapse limit}
As an example of the procedure described above to
generate regular static equilibrium configurations resulting
as final state of a dynamical collapse, we now
consider a well-known static interior for the Schwarzschild metric
which is given by a constant density distribution.
This is the first interior solution obtained by Schwarzschild
himself and it is simple enough to illustrate how a static source
can be achieved from a collapsing scenario as a limit.
Other static interiors, such as those obtained by Tolman
\cite{Tolman},
can be investigated in the similar manner
but we will not go into a detailed analysis for other
spacetimes here.

The main question we address here is: Can we obtain the
above constant density regular Schwarzschild interior equilibrium
configuration as limit of a regular and physically viable
dynamically collapsing matter
cloud, within the framework of general relativity?

In order to have constant density for the interior metric we take
\begin{equation}
    \bar{F}(z)=\frac{\rho_0}{3}z^3 \; ,
\end{equation}
such that
\begin{equation}
    \rho(z)=\rho_0 \; .
\end{equation}
The system of static Einstein equations can easily be
integrated and the pressure is then given by
\begin{equation}\label{staticP}
    p(z)=\rho_0\frac{\sqrt{1-C}-\sqrt{1-C\frac{z^2}{z_b^2}}}
{\sqrt{1-C\frac{z^2}{z_b^2}}-3\sqrt{1-C}} \; ,
\end{equation}
where $C=\frac{2M}{z_b}$ is given by the boundary value
for $\bar{F}(z_b)=2M$ such that it matches a Schwarzschild
manifold with mass parameter $M$ at $p(z_b)=0$.

From the above condition we see that $p(0)=\rho_0\frac{1-\sqrt{1-C}}
{3\sqrt{1-C}-1}$ and we must impose $3\sqrt{1-C}<1$ (corresponding to
the upper mass bound known as Buchdahl limit $M<\frac{4}{9}z_b$
\cite{Buchdahl})
in order for the central
pressure to be finite and positive
\cite{wald}.
Furthermore if we require that the sound speed within the cloud,
defined by $c_s^2=p/\rho$, does not exceed the speed of light we
must require a further constraint on $C$. For example, evaluating
the sound speed at the center we get the condition $C<3/2$
(corresponding to a further upper mass bound $M<\frac{3}{4}z_b$).

The static metric (\ref{metric-eq}) then becomes the well-known
constant density Schwarzschild interior given by,
\begin{equation}\label{static-p(r)}
    ds^2=-\left(\sqrt{1-C\frac{z^2}{z_b^2}}-3\sqrt{1-C}\right)^2dt^2
+\frac{dz^2}{1-C\frac{z^2}{z_b^2}}+z^2d\Omega^2.
\end{equation}
Now we look for the classes of dynamical collapse models
that lead to the above static metric as the limit of collapse.

All those dynamical collapses with $F(r,R)$ going to an
equilibrium configuration limit $F_e(r)=\frac{\rho_0}{3}R_e(r)^3$ and
satisfying the
equilibrium conditions (\ref{ve}) and (\ref{Ge}) will asymptotically
tend to the above static spacetime. This is in general a wide class
of spacetimes, and
not all the possible choices of $F$ will give a physically
valid dynamical collapse evolution.
Of course, in general it will not be possible
to fully integrate Einstein equations during collapse
to give all the global dynamical solutions tending
to the static limit.

Therefore, of the class of all possible $F$ for collapse
we now choose one specific class of models where $p$ does not
vary in time, thus choosing
$p(r,R)=p(r)=p_e(r)$.
The reason to do so resides in the fact that with
this choice we are able to integrate equation (\ref{Gdot}),
thus obtaining an exact analytical solution for collapse
almost fully, while in general exact integration of Einstein
field equations for collapse proves to be unattainable.
From equation (\ref{p}) then we must have,
\begin{equation}
    F(r,R)=y(r)R^3+w(r) ,
\end{equation}
where the freedom to specify $F$ is reflected in the freedom to
choose $y$ and $w$,
and the pressure is then given by
\begin{equation}
    p(r)=-3y(r).
\end{equation}
In such a case, as the cloud shrinks from the initial
radius $R(r,t_i)$ to its final configuration $R_e(r)$, the radial
profile for pressure as a function of the comoving radius $r$
remains unchanged.
From equation (\ref{staticP}) we therefore must take
\begin{equation}
    y(r)=-\frac{\rho_0}{3}\frac{\sqrt{1-C}-\sqrt
{1-C\frac{R_e(r)^2}{z_b^2}}}{\sqrt{1-C\frac{R_e(r)^2}{z_b^2}}-3\sqrt{1-C}}
\end{equation}
and by requiring that the energy density goes to a
constant value $\rho_0$ in the limit of equilibrium we must choose,
\begin{equation}
    w(r)=-\frac{2\rho_0}{3}\frac{\sqrt{1-C}}
{\sqrt{1-C\frac{R_e(r)^2}{z_b^2}}-3\sqrt{1-C}}R_e(r)^3.
\end{equation}
This finally fixes the choice of the free function during collapse.
The boundary of the cloud $r_b$ (chosen such that $R_e(r_b)=z_b$
where the pressure vanishes) corresponds to the physical radius
$R(r_b, t)$ that shrinks from the initial value $R(r_b, t_i)=r_b$ to
the equilibrium value $z_b$. Note that since $p(r_b)=0$ during
collapse, we must have $F(r_b,R_b(t))=w(r_b)=2M$ which means that the
cloud must match to a Schwarzschild exterior.
The energy density during collapse is then given by
\begin{equation}\label{Rho}
    \rho=\frac{y_{,r}R^3+w_{,r}}{R^2R'}-p .
\end{equation}

The above choice of the pressure profile during collapse
was made so that equation (\ref{Gdot}) can be integrated.
Performing the integration we obtain the metric function $G$ as,
\begin{equation}
    G(r,R)=b(r)\left(\frac{y_{,r}r^3+w_{,r}}{y_{,r}R^3+w_{,r}}\right)^2,
\end{equation}
where $b(r)$ is a free function coming from the integration
and we have considered here the initial scaling condition
to be $R(r,t_i)=r$.
Therefore by imposing the equilibrium conditions we must
take the free function $b(r)$ to be
\begin{equation}
    b(r)=\frac{R_e(r)-yR_e(r)^3-w}{R_e(r)}\left(\frac{y_{,r}R_e(r)^3
+w_{,r}}{y_{,r}r^3+w_{,r}}\right)^2.
\end{equation}
This fixes the dynamical evolution completely.
It is easy to see that all quantities obtained depend upon
the unknown function $R(r,t)$, which still needs to be determined.
To obtain the full solution for the dynamical collapsing cloud
one should in principle integrate the two remaining
equations, namely (\ref{phi}) and (\ref{misner}).
Equation (\ref{phi}) can be written as,
\begin{equation}\label{Phi}
  \phi'= \frac{3y_{,r}R^2R'}{y_{,r}R^3+w_{,r}},
\end{equation}
and its integration gives the other metric function
$\phi(r,t)$ also in terms of $R(r,t)$.
The only equation that finally remains to be solved
then is the equation of motion
(\ref{misner}) which can be written as,
\begin{equation}
  \dot{R} = -e^{\phi}\sqrt{\frac{F}{R}+G-1}.
\end{equation}
We can treat this equation as an ODE for the function $R$
of $t$ for each fixed radial shell $r$.
We then note that this is a first order ODE in $t$ that,
given the continuity of the functions involved, always
admits a solution. Therefore integration
of this equation gives the metric function $R(r,t)$
and thus solves the system of Einstein equations
entirely.

The dynamical spacetime (\ref{metric}) tending to the
static limit given by the metric (\ref{static-p(r)}) is then
finally written as,
\begin{equation}
    ds^2=-e^{2\int_0^r\frac{3y_{,r}R^2R'}{y_{,r}R^3+w_{,r}}d\tilde{r}}dt^2+
    \frac{R'^2dr^2}{b(r)\left(\frac{y_{,r}r^3+w_{,r}}{y_{,r}R^3+w_{,r}}
\right)^2}+R^2d\Omega^2 \; .
\end{equation}

Equation (\ref{Rho}) can be used to study the behaviour of
the central density during collapse. If we suppose that
all the functions involved are finite and continuous in the
radial direction, we can then expand the area coordinate
$R(r,t)$ near the center as
\begin{equation}
    R(r,t)=a_1(t)r+a_2(t)r^2+...
\end{equation}
with the initial conditions $a_1(t_i)=1$ and $a_j(t_i)=0$
for $j>1$. Then as collapse evolves $R(r,t)$ will tend to
the limit $R(r,t)\longrightarrow R_e(r)$ with
$R_e(r)=a_{1e}r+a_{2e}r^2+...$ from which we can see that
the central
pressure will have the following behaviour
\begin{equation}
    \rho(0,t)=\rho_0+\rho_0\frac{2\sqrt{1-C}}{1-3\sqrt{1-C}}
\left(1-\frac{a_{1e}^3}{a_1(t)^3}\right).
\end{equation}
From the above equation for the central density we obtain
a series of constraints that can be imposed on the model
in order for it to be physically valid:
\begin{itemize}
  \item Positivity of initial density:
  $$
  \frac{2\sqrt{1-C}}{1-3\sqrt{1-C}}(1-a_{1e}^3)>-1.
  $$
  \item Initial density smaller than final density:
  $$
  \frac{2\sqrt{1-C}}{1-3\sqrt{1-C}}(1-a_{1e}^3)<0.
  $$
  \item Initial sound speed smaller than $1$:
  $$
  \frac{a_{1e}^3\sqrt{1-C}}{1-\sqrt{1-C}-2a_{1e}^3\sqrt{1-C}}<1.
  $$
\end{itemize}
Note that the last constraint above implies that imposing
the speed of sound within the
cloud to be smaller than 1 at the initial time is sufficient
for the same to be smaller than 1 during the whole evolution.

Further to this, one would like to require that there
are no shell-crossings during the collapse evolution. This
means that the shell labeled by $r_1$ and $r_2$ do not
overlap at any point during the evolution. The singularities
that originate due to shell crossings (namely when $R'=0$)
signal a breakdown of the coordinates at that point and
therefore imply the impossibility to describe the future evolution
of the cloud using that coordinate system. These are generally
believed to be weak singularities that have no physical relevance
being due only to the coordinate choice
\cite{cross}.
The requirement that
there are no shell-crossing singularities is given by $R'>0$
which, by making use of equations (\ref{rho}) and equation
(\ref{phi}), can be written as,
\begin{equation}
R' = \frac{F_{,r}}{(\rho+p)R^2}>0.
\end{equation}
It follows then that provided that the weak energy condition
is satisfied throughout collapse, the condition for avoidance of
shell-crossing is just given by $F_{,r}>0$. Then it is easy to check
for the above toy model that if the condition is satisfied at
equilibrium it will be satisfied throughout collapse.
This may be considered as a very reasonable physical
condition in that it is basically the requirement that the
mass should be increasing with the increasing radial
coordinate $r$.

We note that imposing during collapse physically realistic
conditions
such as regularity of initial data, avoidance of shell crossing,
sound speed smaller than the speed of light, allows us to restrict
the array of possible scenarios that can evolve from collapse. This
eventually allows us to narrow down the large number of existing
static interior solutions on the basis of those that can be
achieved via realistic collapse.

\section{Concluding remarks}
It is clear for many years now, that the physics of
gravity and the general theory of relativity must play
a crucial role when it comes to understanding very
compact objects in nature. This has
been the key motivation for the many attempts over past
years for a fully general relativistic modeling and
description for such objects. These models also greatly
help in understanding the accretion disk properties and
related observational details for such objects and the
associated high energy phenomena, with observations
now offering an increasing degree of preciseness
and detail.

Since the original interior solution with constant
density was provided by Schwarzschild in 1915, there have
been a lot of studies on matter sources in general relativity.
Static interiors sustained by perfect fluid pressures were
also investigated by Tolman
\cite{Tolman}
who found a number
of solutions describing compact objects. Interiors sustained
by anisotropic pressures were investigated in
\cite{Bowers} and \cite{deleon},
while interiors sustained only by tangential pressures were firstly studied in
\cite{florides}.
Since then the zoo of static interiors to
the Schwarzschild vacuum spacetime has become very crowded
(see \cite{DL}
for an overview of interior
solutions with perfect fluids). In principle, given a generating
function in the form of $\phi_{,z}$, it is possible to construct
any interior matching smoothly to the vacuum Schwarzschild metric
\cite{Visser}.
Nevertheless, as pointed out by Lake and Delgaty \cite{DL},
most known solutions fail to
fulfill some of the physical requirements listed above.
Furthermore, such solutions need not arise naturally from
a physically realistic collapsing configuration and their
connection with collapsing models has not been thoroughly
investigated.
From the point of view of astrophysics the critical densities that
lead to the formation of different kinds of compact objects were
first studied in
\cite{wheeler}
and have been thoroughly investigated ever since.
Nevertheless our knowledge of the states of matter that are possible
in extreme conditions is still very limited and it is possible that,
together with the neutron star matter equation of state, other kinds
of `exotic' fluids might exist in nature, possibly for more compact and
dense objects.
For this reason many theoretical models for `exotic' compact stars
have been proposed over the years.
What we really need to decide is which ones
will be the most relevant models to choose from, from a
physical perspective, out of such a wide collection and variety
offered by general relativity.

Under the situation, while a general relativistic description
of such static compact objects where strong gravity effects are
highly important is not only essential but is also inevitable,
what is really necessary is to isolate the models which
would be physically most relevant and significant
and the equations of general relativity must necessarily be the starting
point in this search.
It is in this spirit that we argued here that the static configurations
that arise from a general relativistic gravitational collapse,
which we already know to be a very relevant physical process
in the universe, could provide us with a physically preferred
set of models, as opposed to the plethora of very many other
static solutions to Einstein equations as mentioned above.
For example, an important question that needs to be addressed is
whether the gravastar models described in
\cite{gravastar2}
can be obtained via a collapse mechanism such as the one described here.
In such a case negative pressures and non-ideal fluid models are likely
to play a significant role as well. Also modifications to general relativity
in the strong field may be required to balance the gravitational attraction.
A simple and generic toy model for relativistic collapse leading
to the formation of such exotic compact objects is still missing.
Such a model would play a role similar to that of Oppenheimer-Snyder
collapse for black hole physics
\cite{OS}.

In other
words, we propose here that the compact static configurations
as arising from gravitational collapse, as per the procedure
outlined here, may play a physically more significant and
essential role, and that these should be investigated in
further detail.
Further work in that direction in under way.
We showed here how it is possible to obtain
perfect fluid static interiors from dynamical collapse
with regular initial data. This shows that analytic models
that have been considered to describe the equation of state
for neutron stars, such as the Tolman interiors
(see for example
\cite{Lattimer}),
can be obtained quite naturally from gravitational collapse.
Further physical considerations are needed if one
wishes to apply similar models in realistic astrophysical
scenarios, and the study of different profiles for densities
and pressures will also be important in this connection.

\label{lastpage}

\end{document}